\documentclass[fleqn,usenatbib]{mnras}
\usepackage[T1]{fontenc}
\DeclareRobustCommand{\VAN}[3]{#2}
\let\VANthebibliography\thebibliography
\def\thebibliography{\DeclareRobustCommand{\VAN}[3]{##3}\VANthebibliography}

\usepackage{graphicx}	
\usepackage{amsmath}	
\usepackage{amssymb}	
\usepackage{bm}		
\usepackage{pdflscape}	
\usepackage{supertabular, multicol} 
\usepackage{ctable} 
\usepackage{longtable} 
\usepackage{hyperref}
\usepackage{newtxtext,newtxmath}

\newcommand{\photu}{photon units}

\newcommand{\galex}{{\it GALEX}}

\title[UV Variations]{Time variations in the UV background}

\author[J. Murthy et al.]{
Jayant Murthy$^{1}$\thanks{E-mail: jmurthy@yahoo.com}, Amith S. Gowtham\thanks{E-mail: amithsgowtham@gmail.com} \\
$^{1}$Indian Institute of Astrophysics, Koramangala 2nd block, Bangalore, 560034, India
}

\pubyear{2022}

\begin{document}
\maketitle

\begin{abstract}
We have found variations in the diffuse ultraviolet background on the scale about 10 days over the 10 year life of the \galex\ mission. These variations are only apparent in the near-ultraviolet band of \galex\ and are most likely related to variations in the zodiacal light due to brightness variations of the solar NUV flux. The variations can be as high as 200 \photu\ in the NUV, amounting to about 15\% of the zodiacal light or $<10\% $ of the total NUV diffuse radiation. There is no effect on the far-ultraviolet band of \galex . Further work will require better observations chosen specifically for the purpose of investigating this component and will be difficult to obtain.

\end{abstract}




\section{Introduction}

The first observations of the ultraviolet (UV) background were made by \citet{Hayakawa1969} followed by a number of (mostly) spectroscopic observations in the spectral range between 1200 -- 3000 \AA . These early observations have been reviewed by \citet{Bowyer1991} and \citet{Henry1991} and showed that there was a Galactic component increasing toward the Galactic plane and an extragalactic component observable at high Galactic latitudes, where the extinction from interstellar dust is low. The first all-sky observations came with the flight of the \galex\ ({\it Galaxy Evolution Explorer}) imaging mission \citep{Martin2005, Morrissey2007} and the {\it SPEAR} (Spectroscopy of Plasma Evolution from Astrophysical Radiation) spectroscopic mission \citep{Edelstein2006, Edelstein2006b}, both of which mapped the diffuse background over the entire sky  \citep{Seon2011, Hamden2013, Murthy2014apss, Murthy2014apj}.

The diffuse ultraviolet background is comprised of several components, which are difficult to separate in imaging, or even spectroscopic, observations. These components \citep{Leinert1998,Murthyreview2009} include:
\begin{itemize}
    \item Airglow: Emission lines arising in the Earth's atmosphere.
    \item Zodiacal light (ZL): Sunlight scattered from interplanetary grains. The zodiacal light is on the same order as the DGL at wavelengths longer than 2000 \AA\ with no contribution at shorter wavelengths.
    \item Diffuse Galactic light (DGL): Predominantly starlight scattered by interstellar grains with contributions from molecular hydrogen (H$_{2}$) fluorescence \citep{Hurwitz1994}, emission lines from ionized gas \citep{Jakobsen1981}, or two-photon recombination radiation \citep{Reynolds1990}.
    \item Extragalactic background light (EBL):  Predominantly due to the integrated light of galaxies  \citep{Xu2005,Driver2016} with smaller contributions from the integrated light of QSOs \citep{Madau1992}) and the IGM \citep{Martin1991}).
\end{itemize}    

The all-sky nature of the \galex\ and the SPEAR observations have allowed us to separate the components based on their distribution over the sky. The foreground emission (airglow and zodiacal light) is dependent on the observation time and date and \citet{Murthy2014apss} has characterized their contribution to the total diffuse background in the \galex\ data while \citet{Akshaya2018, Akshaya2019} have mapped the contributions from the Galactic and extragalactic components at high latitudes. Similarly, \citet{Seon2011} and \citet{Jo2017} have mapped the diffuse emission observed by SPEAR, both the continuum dust emission and the molecular hydrogen fluorescence with the advantage over \galex\ of spectroscopy but a poorer sensitivity and spatial resolution.

Of all the contributors to the diffuse observations, the only one which might be expected to be time-variable is the airglow, which is a function of the local time of observation \citep{Murthy2014apss}, and it was a surprise when \citet{Akshaya2018} found time-variable variations in the \galex\ observations in Virgo, which they attributed to unknown atmospheric sources. \galex\ was never intended to probe variability in the DGL and it is difficult now, after the mission has ended, to divine the source of the variability. Nevertheless, it is important to understand the impact of the variability on the diffuse background. We have gone through the \galex\ database and found three locations with a sufficient number of observations to check for variability. 

\section{Observations and Data Analysis}

The \galex\ spacecraft and its mission have been described by \citet{Martin2005} and \citet{Morrissey2007}. \galex\ included two photon-counting detectors: the far-ultraviolet (FUV: 1344 -- 1786 \AA\ with an effective wavelength of 1539 \AA) and the near-ultraviolet (NUV: 1771 -- 2831 \AA\ with an effective wavelength of 2316 \AA). The NUV instrument observed the sky for the entire length of the mission from 2003 June 7 until 2013 June 28 while the FUV detector failed permanently in 2009 May, with intermittent interruptions even before that date. The field of view (FOV) of the instrument was $1.25^{\circ}$ with a spatial resolution of 5 -- 10\arcsec. 

Most of the sky was observed as part of the All-sky Imaging Survey (AIS) with a typical exposure time of about 100 seconds with a few locations targeted for deeper observations of up to 100,000 seconds. These longer observations were comprised of a series of visits, each of less than 1000 seconds in length, taken during local (spacecraft) night and sometimes, but not always, taken in consecutive orbits. \citet{Murthy2014apj} used the \galex\ data to map the diffuse background over the sky at a spatial resolution of $2'$. Because the foreground emission (airglow and zodiacal light) was different for each visit \citep{Murthy2014apss}, they created files of the sky background with the foreground emission subtracted for each visit \footnote{\url{https://archive.stsci.edu/prepds/uv-bkgd/}}. We have binned the data in those files into $6'$ bins and used those to study the time dependence of the diffuse UV background.

\begin{table*}
	\centering
	\caption{\galex\ Observations}
	\label{tab:ob_count}
	\begin{tabular}{rrrrrrrrrr}
		Field & Field ID$^{a}$ & $l$  & $b$ & N(NUV)$^{b}$ & N(FUV)$^{b}$ & NUV$^{c}$ & FUV$^{c}$ & $\Delta NUV^{d}$ & $\Delta FUV^{d}$\\
		\hline
		1. & WDST\_LB\_227 & 176.1  &   -24.7  & 872 & 138 &   1910    &    2440  & 50.8 & 40.7  \\
		2. & WDST\_LDS\_749B & 54.3  &   -34.9  & 1388 & 458 &  850    &    670  & 35.0 & 19.4 \\
		3. & Virgo Cluster & 284.7  &    74.4  &   1143 & 12& 670    &    440  & 18.1 & 21.4    \\
		\hline
		\multicolumn{5}{l}{$^{a}$From \galex\ observation identifier.}\\
		\multicolumn{5}{l}{$^{b}$Number of visits.}\\
		\multicolumn{5}{l}{$^{c}$Mean value of DGL in \photu .}\\
		\multicolumn{5}{l}{$^{d}$Mean value of standard deviation of DGL in \photu .}\\
	\end{tabular}
\end{table*}

\begin{figure}
    \centering
    \includegraphics[width=3in]{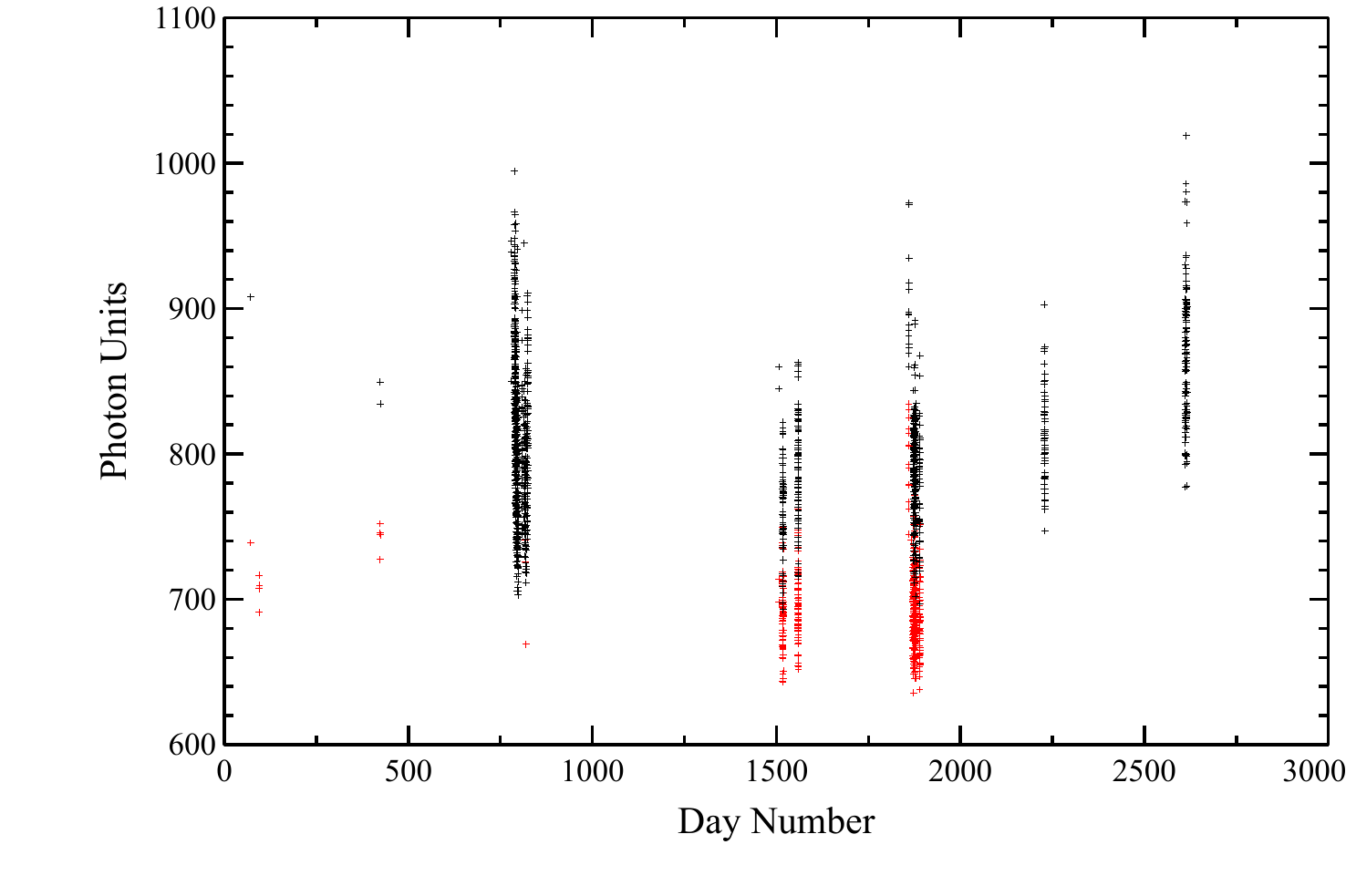}
    \caption{Time history of a single $0.1^{\circ} \times 0.1^{\circ}$ bin with FUV shown in red. The day number is counted from 2003, June 10, approximately when \galex\ observations began.}
    \label{fig:timehistory}
\end{figure}

 Although \galex\ observed a significant fraction of the sky, most areas were covered by fewer than 5 visits and there were only 3 regions in which there were more than 100 visits in the NUV (Table \ref{tab:ob_count}). the first two were repeat calibration observations of white dwarfs with the third being a set of deep observations of Virgo \citep{Boissier2015}. For reference, we have given the mean values of the diffuse background in those fields as well as the standard deviation of the values in each bin. It is interesting to note, but out of the scope of this work, that the standard deviations are higher at low Galactic latitudes, perhaps because of the complexity of the ISM near the Galactic Plane.
 
 As discussed above, we broke the sky into $6'$ bins and extracted the diffuse surface brightness in each bin as a function of day number, starting on 2003, June 10 (Julian day 2452800.5). The airglow increases with time from local midnight and we have only used observations which fell two hours of orbital midnight, where the contribution from airglow is less than 5 \photu\ \citep{Murthy2014apss}. Because there was no intention of measuring the time variation of the diffuse background, the data were taken at random intervals over the course of the mission, sometimes in a series of groups (Fig. \ref{fig:timehistory}). It is not readily apparent in this plot but we do see variations both on the scale of a few days and over the entire set of observations.
 
 \begin{figure}
    \centering
    \includegraphics[width=3in]{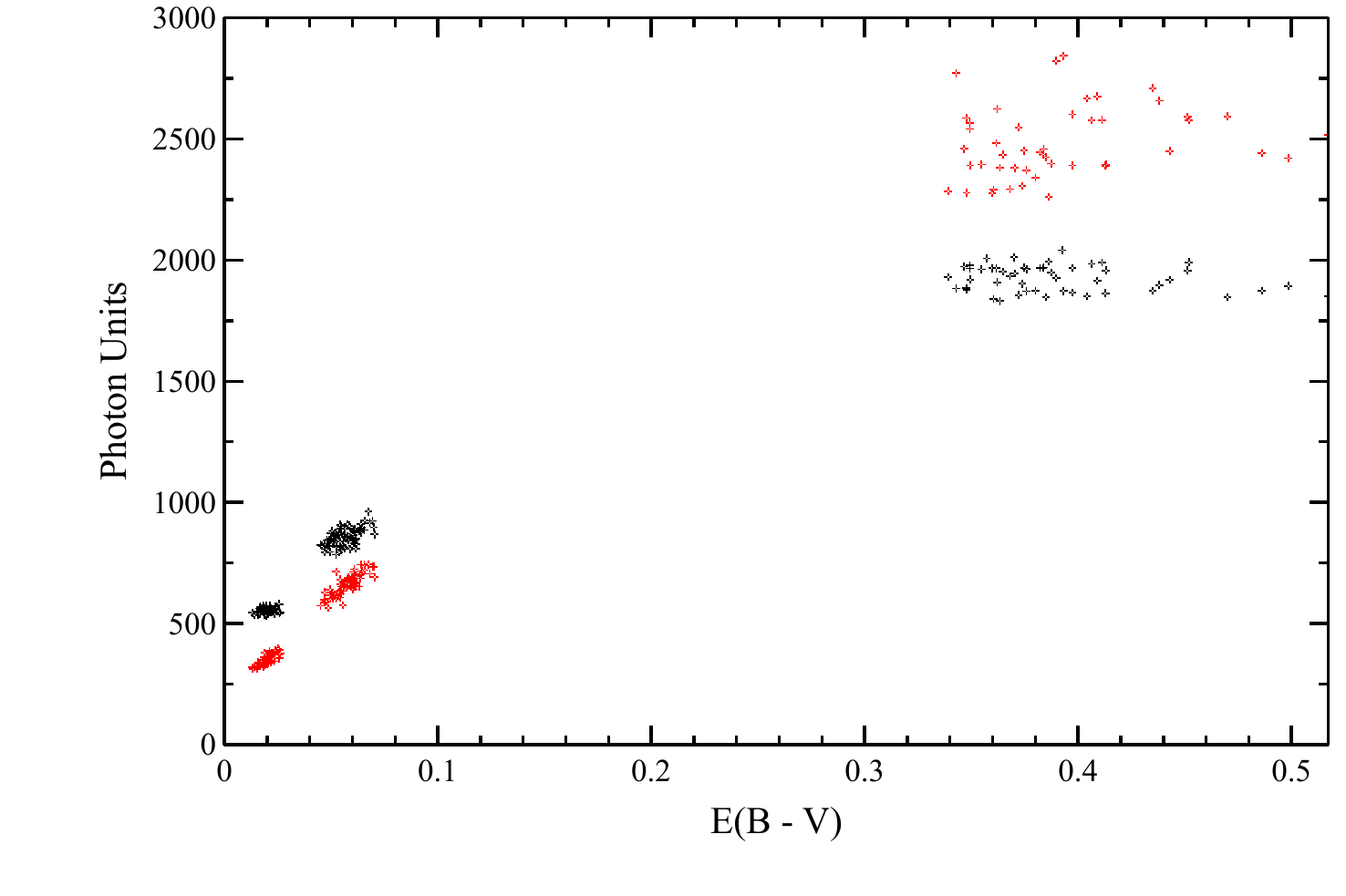}
    \caption{Time-averaged FUV (red) and NUV surface brightness as a function of the Planck E(B - V).}
    \label{fig:ebv_means}
\end{figure}

\begin{table}
	\centering
	\caption{UV-Reddening Relations}
	\label{tab:ebv_means}
	\begin{tabular}{lll}
		Band & Offset$^{a}$ & Slope$^{b}$\\
		\hline
		FUV & 190 & 8105  \\
		NUV & 405 & 7894 \\
\hline
		\multicolumn{3}{l}{$^{a}$\photu .}\\
		\multicolumn{3}{l}{$^{b}$\photu\ mag$^{-1}$ .}\\
	\end{tabular}
\end{table}

Without loss of generality, we can break up the observed signal in each pixel into a constant component, estimated from the mean of the surface brightness over time, and a time-variable component, given by the deviations from the mean. The mean value for each pixel is representative of the DGL in that location and we have plotted it as a function of the Planck E(B - V) \citep{PlanckDust2016} for each of the three regions in Fig. \ref{fig:ebv_means}. The UV emission arises within a few hundred parsecs of the Earth \citep{Murthy_dustmodel2016} and is linearly correlated (Table \ref{tab:ebv_means}) with the reddening until an E(B - V) of about 0.15, when the UV saturates. The modeling of each region is dependent on local conditions \citep{Murthy_sahnow2004} and we will leave a discussion of the DGL to a future work.

\begin{figure}
    \centering
    \includegraphics[width=3in]{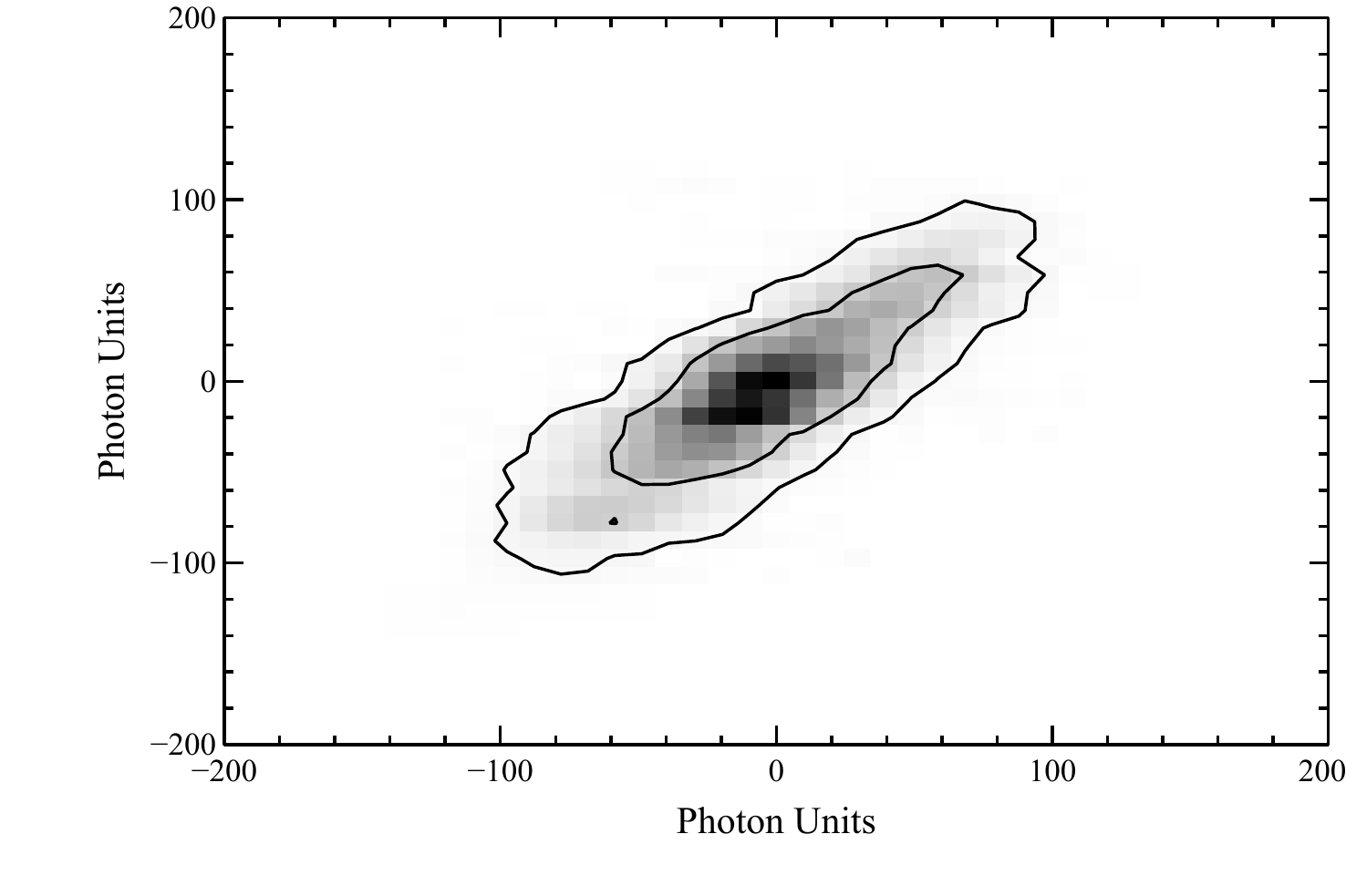}
    \caption{Correlation between different pixels in the NUV after subtracting the mean. Contours represent 70\% and 95\% of the data points.}
    \label{fig:xycorr}
\end{figure}

\begin{figure}
    \centering
    \includegraphics[width=3in]{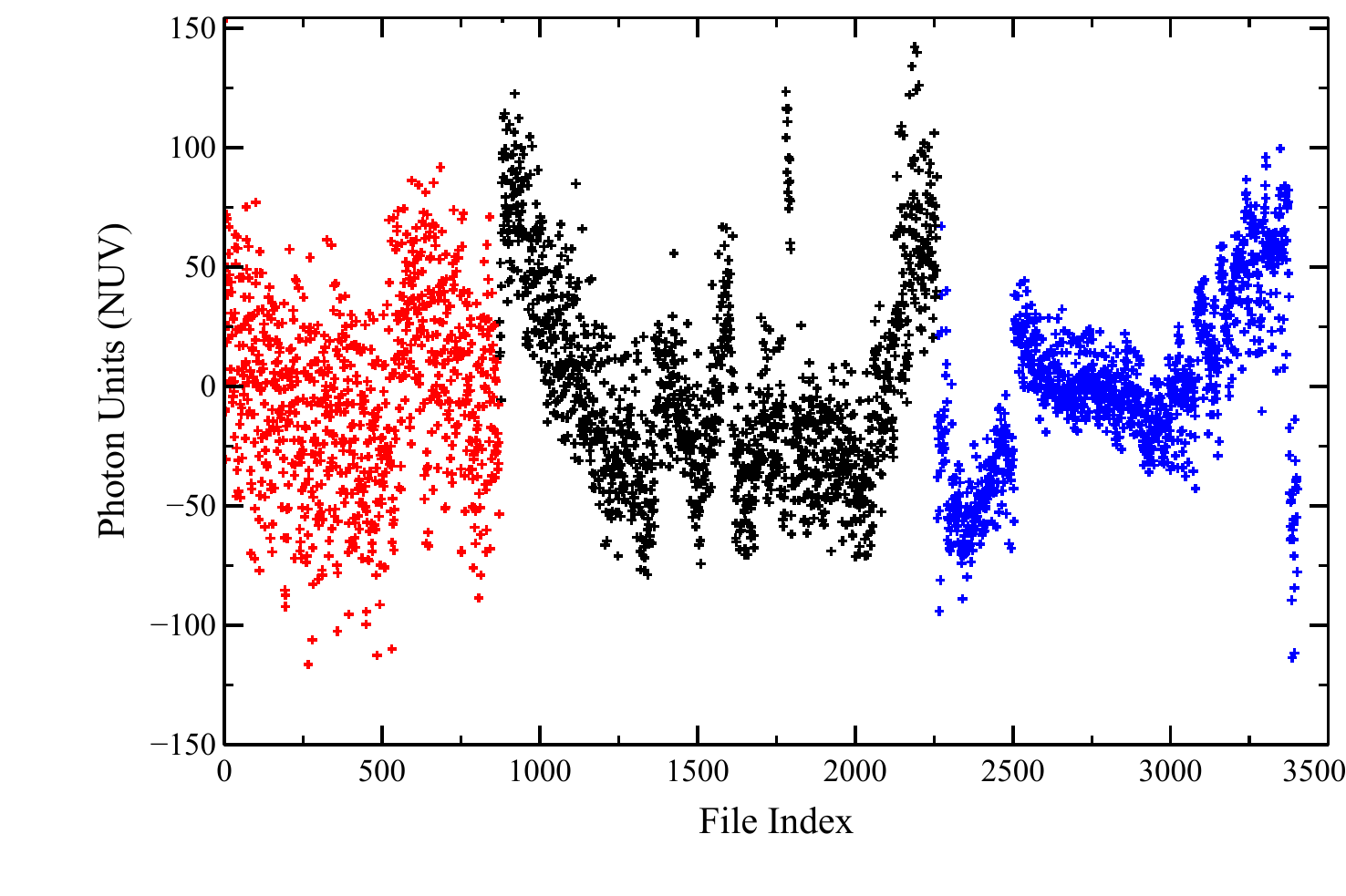}
    \caption{Time variable component of NUV surface brightness per pixel plotted as a function of file index for Field 1 (red), Field 2 (black), and Field 3 (blue) with one point per visit. Note that the time gap between visits is not uniform.}
    \label{fig:nuv_final_index}
\end{figure}

\begin{figure}
    \centering
    \includegraphics[width=3in]{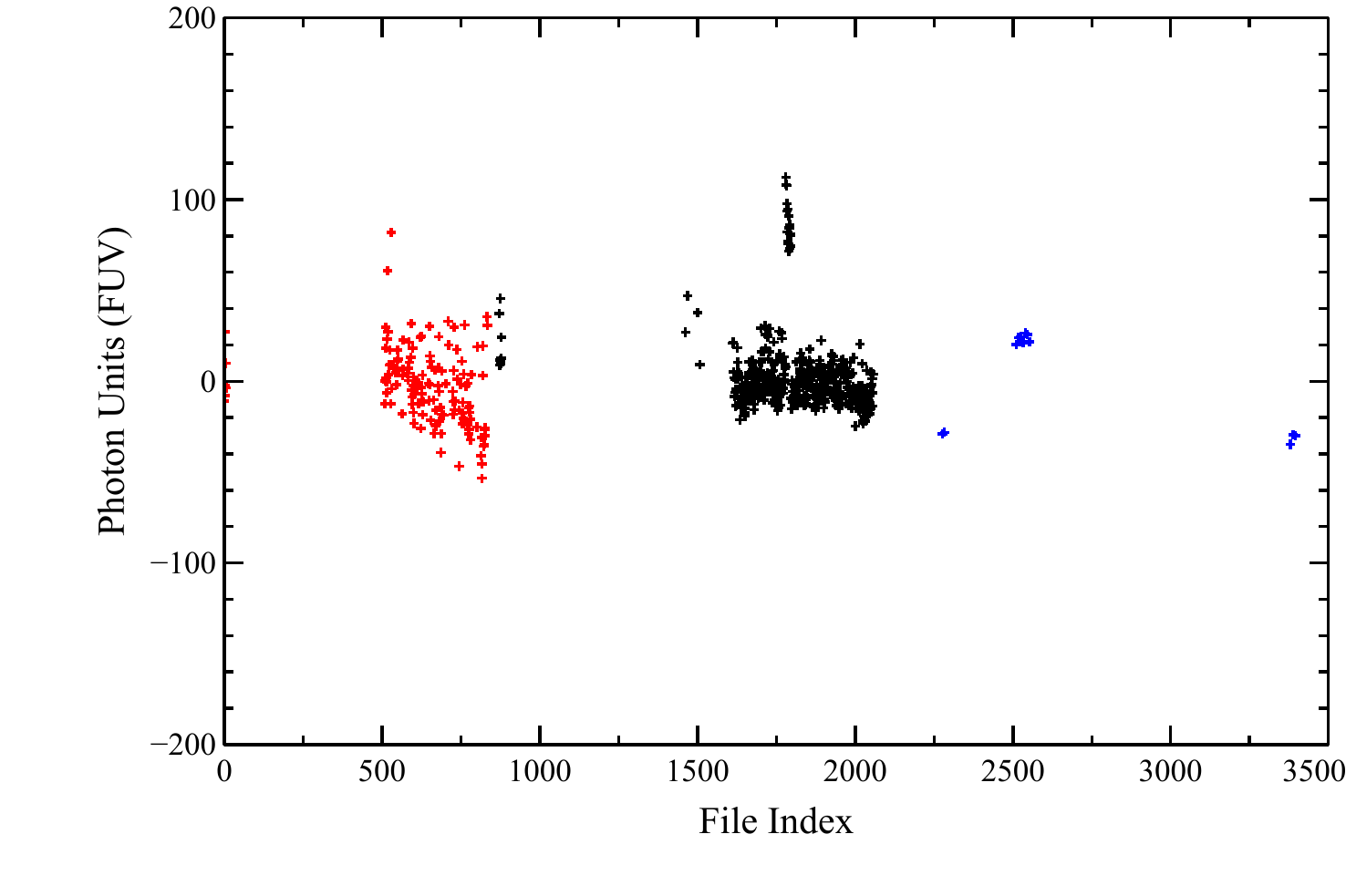}
    \caption{Time variable component of FUV surface brightness per pixel plotted as a function of file index for Field 1 (red), Field 2 (black), and Field 3 (blue) with one point per visit. There are many fewer points than the NUV because of the intermittent failures of the FUV HVPS.}
    \label{fig:fuv_final_index}
\end{figure}

\begin{figure*}
    \centering
    \includegraphics[width=6in]{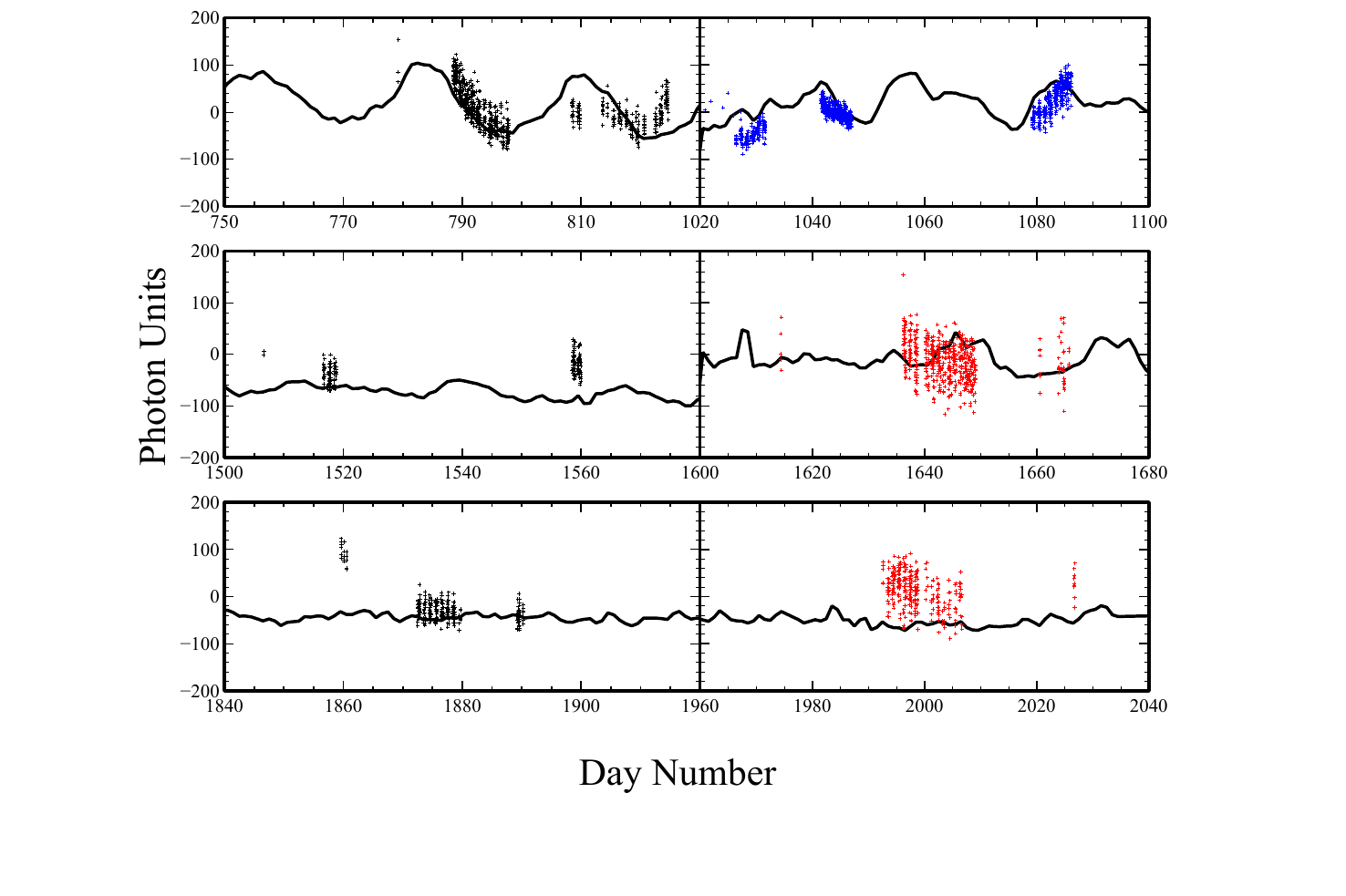}
    \caption{Variations in the NUV surface brightness as a function of the day number. The y-axis is identical in each plot with the day number (x-axis) in groups of 80 days. The colors are the same as in Fig. \ref{fig:nuv_final_index}. The solid line in each panel shows the variations in the solar irradiance, with an arbitrary scale and offset.} 
    \label{fig:nuv_mjd}
\end{figure*}

\begin{figure}
    \centering
    \includegraphics[width=3in]{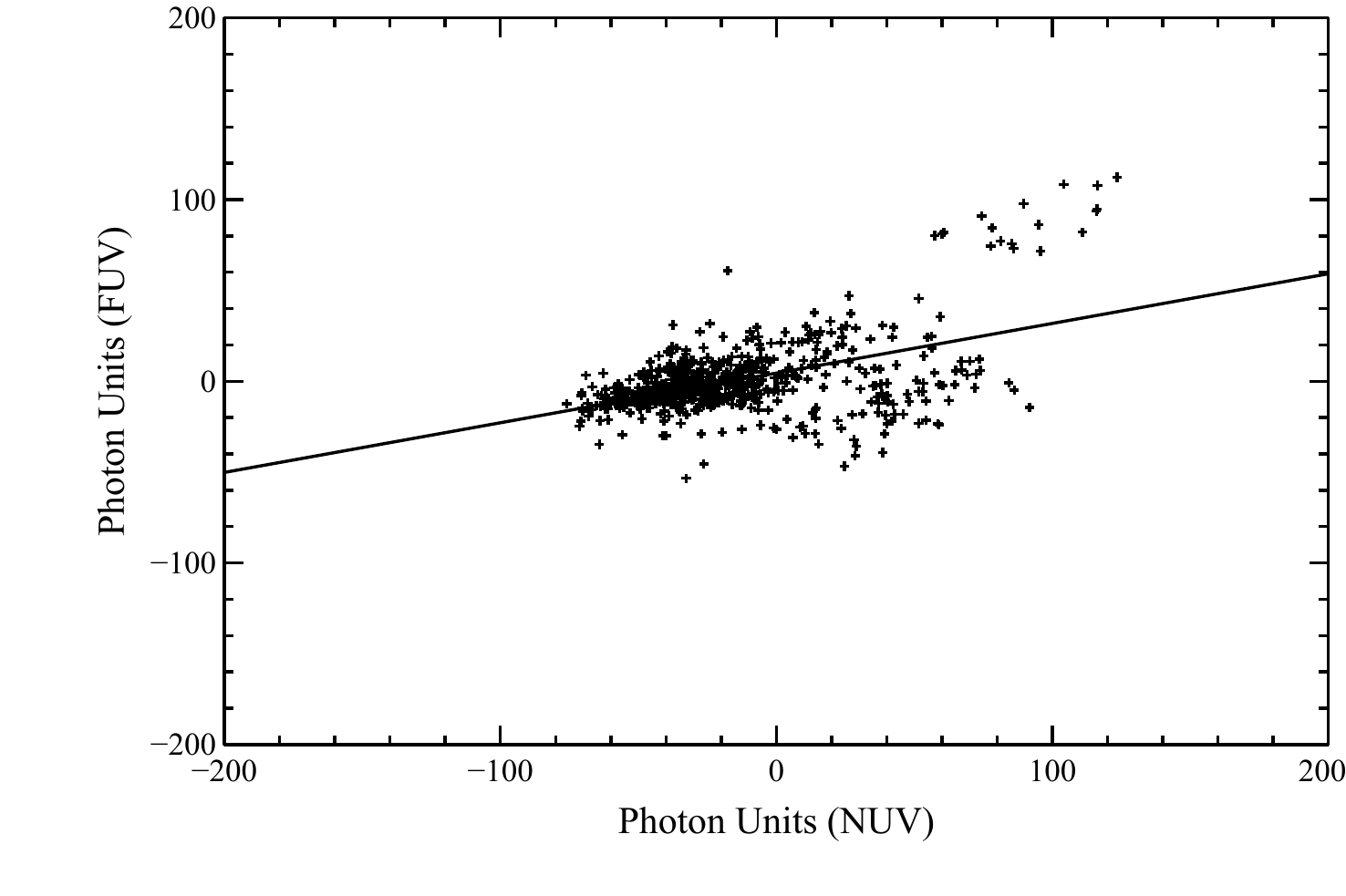}
    \caption{Variations in the FUV surface brightness plotted against the NUV surface brightness. The line is the best fit: FUV $= 4.5 + 0.27 \times\ NUV$ with a correlation coefficient of 0.5.}
    \label{fig:fuv_nuv_corr}
\end{figure}

 The deviations from the mean are due to the time-variable part of the diffuse background, whatever its origin may be. As a reminder, a \galex\ observation extends over a $1^{\circ}$ field and we found, empirically, that the time variations in each pixel are correlated in a given visit in the NUV (Fig. \ref{fig:xycorr}); ie., the change in the observed signal is due to a general increase in the background over the entirety of the \galex\ field. We therefore averaged the surface brightness over a given visit and plotted the values as a function of visit number in Fig. \ref{fig:nuv_final_index} (NUV) and \ref{fig:fuv_final_index} (FUV) for each of the three fields in Table \ref{tab:ob_count}. These visits were spread over a decade but there were several periods where many observations were made over the period of a few months (Fig. \ref{fig:nuv_mjd}). There was much more scatter in the first two fields (Table \ref{tab:ob_count}), which were at moderate galactic latitudes, but the trend of coordinated variability is clear in all three fields. This is particularly striking in Field 3 (Virgo) where the observations cover an area of approximately $5^{\circ} \times 3^{\circ}$. There are not enough FUV observations to discern any long term trends but we do see a mild correlation between the FUV and the NUV (Fig. \ref{fig:fuv_nuv_corr}), with the FUV variations at a significantly lower level. There is a spike in both the NUV and FUV around 13 July, 2008 (day 1860) which is, again, due to a general increase in the background level. We do not have any way of reconstructing what happened on that date and have left those points out of our analysis.
 
\section{Discussion}

There are several possibilities for the origin of the observed variability. The spatial scale of the variability over three different regions over several square degrees suggests that the source must be local. We will consider the three possibilities below. The first is that there is an increase in the dark current due to an increase in the radiation environment of the spacecraft. We checked for any correlation with the spacecraft position, whether near the South Atlantic Anomaly or at high latitudes (note that the inclination of the orbit was $29^{\circ}$). Moreover, the two detectors are identical, except for the photocathode, and one would expect the count rate to be identical in both detectors. This would manifest as a higher background (in \photu )in the FUV than the NUV, due to the conversion from counts to \photu . Thus, with the exception of the isolated event around 13 July, 2008, the variability is not due to increases in the dark count.

We know that airglow does vary over the course of an observation but \citet{Murthy2014apss} have shown that there is little airglow contribution within two hours of orbital midnight. The primary emission lines in the \galex\ bands are the geocoronal O I lines at 1304 and 1356 \AA\ in the FUV and at 2471 \AA\ in the NUV \citep{Morrison1992, feldman_OI_1992} with \citet{kulkarni_fuv_2021} suggesting that continuum two-photon emission from the atmosphere could contribute about 20 \photu\ to the background but all these mechanisms would contribute more to the FUV emission than the NUV, which is not what we see.

In our estimation, the fact that we see a variation in the NUV surface brightness and not in the FUV points to an origin in the zodiacal light \citep{Murthy2014apss}. The zodiacal light is due to scattering of the solar photons by interplanetary dust and variations might be due to either changes in the dust distribution \citep{kelsall_zodi_1998} or to a change in the solar flux. We have plotted the solar irradiance from \citet{Woods_solar_2018} in Fig. \ref{fig:nuv_mjd}, with an arbitrary (but constant) scale and a different offset in each panel. A detailed analysis would involve modeling of the interplanetary dust distribution and is beyond the scope of this work.

\section{Conclusions}

We have found a time-variable component to the diffuse background in the \galex\ NUV band with a maximum variation of 100 \photu\ from the mean , with no corresponding variation in the \galex\ FUV observations. We believe that the most likely source of this variation is the zodiacal light and the time variability of the solar NUV flux, both of which show a similarity in shape. Unfortunately, the \galex\ data are sparse and are irregularly distributed in space and time and we cannot conclusively prove the origin of this component. In practice, we can probably ignore the time-variable component to the diffuse UV background and the FUV data are unaffected.

\section*{Acknowledgements}

This research has made use of NASA's Astrophysics Data System Bibliographic Services. We have used the GnuDataLanguage (http://gnudatalanguage.sourceforge.net/index.php) for the analysis of this data. The data presented in this paper were obtained from the Mikulski Archive for Space Telescopes (MAST). STScI is operated by the Association of Universities for Research in Astronomy, Inc., under NASA contract NAS5-26555. Support for MAST for non-HST data is provided by the NASA Office of Space Science via grant NNX09AF08G and by other grants and contracts.

\section*{Data Availability}

The processed time-series data from the three fields underlying this article will be made available to the reader upon request from the authors.



\bibliographystyle{mnras}
\bibliography{murthy}

\end{document}